\documentclass[10pt,final, twocolumn]{IEEEtran}
\usepackage{graphicx,cite,epsfig,amssymb,amsmath}

\newtheorem{assumption}{Assumption}

\begin{document}
\title{A Dual-Directional Path-loss Model in 5G Wireless Fractal Small Cell Networks}
\author{\IEEEauthorblockN{Jiaqi Chen\IEEEauthorrefmark{1},
Fen Bin\IEEEauthorrefmark{1},
Xiaohu Ge\IEEEauthorrefmark{1},
Qiang Li\IEEEauthorrefmark{1},
Cheng-Xiang Wang\IEEEauthorrefmark{2}\\
}
\IEEEauthorblockA{\IEEEauthorrefmark{1} School of Electronic Information \& Communications\\
Huazhong University of Science \& Technology.\\
}
\IEEEauthorblockA{\IEEEauthorrefmark{2} School of Engineering \& Physical Sciences\\
Heriot-Watt University.\\
Corresponding Author: Xiaohu Ge, Email: xhge@mail.hust.edu.cn}\\
\thanks{\scriptsize{This paper has submitted to IEEE ICC 2017.}}}
\maketitle

\begin{abstract}
With the anticipated increase in the number of low power base stations (BSs) deployed in small cell networks, blockage effects becoming more sensitive on wireless transmissions over high spectrums, variable propagation fading scenarios make it hard to describe coverage of small cell networks. In this paper, we propose a dual-directional path loss model cooperating with Line-of-Sight (LoS) and Non-Line-of-Sight (NLoS) transmissions for the fifth generation (5G) fractal small cell networks. Based on the proposed path loss model, a LoS transmission probability is derived as a function of the coordinate azimuth of the BS and the distance between the mobile user (MU) and the BS. Moreover, the coverage probability and the average achievable rate are analyzed for 5G fractal small cell networks. Numerical results imply that the minimum intensity of blockages and the maximum intensity of BSs can not guarantee the maximum average achievable rate in 5G fractal small cell networks. Our results explore the relationship between the anisotropic path loss fading and the small cell coverage in 5G fractal small cell networks.
\end{abstract}

\IEEEpeerreviewmaketitle

\section{Introduction}
It is now widely accepted that the fifth generation (5G) cellular networks will have significant changes that include low-power base stations (BSs) and dense deployments, higher spectrums, massive bandwidths and large numbers of antennas \cite{R1,R2,X2}. With the anticipated increase in the number of low-power BSs deployed in small cell networks, variable propagation losses make it hard to predict coverage of small cell networks. Moreover, blockage effects become more sensitive for wireless communication systems with higher spectrums \cite{R3} such that the user association and the system performance become more seriously. In general, blockage effects is contained by a shadowing fading modeled as a log-normal distributed random variable with a uniform path loss exponent. However, different path loss exponents of special environments \cite{R4,X1} and distance-dependence of blockage effects \cite{R5} are ignored in the traditional performance analysis of cellular networks.

Due to the deployment of a variety of low-power BSs adopting millimeter wave (mmWave) technologies, the coverage of small cell networks is less regular. In 5G small cell networks, wireless transmissions are often divided into Line-of-Sight (LoS) and Non-Line-of-Sight (NLoS) wireless transmissions. Up to now, a simplistic path loss model with a fixed path loss exponent in the entire plane is widely considered in previous literatures \cite{R6}\cite{R7}\cite{R8}. These studies focused on random heterogeneous cellular networks, but the difference of LoS and NLoS wireless transmissions is ignored in cellular networks. The difference between LoS and NLoS wireless transmissions was investigated by path loss models in \cite{R5}\cite{R9,R10,R11}. A path loss model incorporated blockage effects was proposed in \cite{R5}, which matched experimental trends that the distribution of the number of blockages in a link was proved to be a Poisson random variable with parameters depending on the length of links. Based on the results in \cite{R5}, different LoS and NLoS path loss laws with distance dependent LOS probability functions were used to analyze the mmWave performance of coverages and transmission rates \cite{R9}. A multi-slope path loss models were proposed in \cite{R10}, where different distance ranges were subject to different path loss exponents. The impact of a sophisticated piece-wise path loss model with probabilistic LoS and NLoS transmissions on the performance of dense small cell networks was investigated in \cite{R11}, where the LoS probability was assumed as a linear function with the distance between a mobile user (MU) and a BS.

It was proved that the practical coverage of cellular networks has the fractal characteristic due to anisotropic path loss exponents \cite{R17}. Measurement results of path loss exponent in \cite{R12} indicated that the path loss exponent varies with the angles of the receiver relative to the BS. In fact, the probability of LoS transmission in a link is associated with the distance of the link and the direction of the link relative to the BS. In this paper, based on the fractal characteristics of 5G wireless small cell networks, a dual-directional path loss model is proposed in which the probability of LoS transmission is derived by the coordinate azimuth of a BS and the distance between a MU and the BS. Moreover, all studies in \cite{R5}\cite{R9,R10,R11} assumed that a user is associated with the nearest BS, or the nearest visible BS. Considering the random blockage effect and the large number of low-power BSs, the channel state of the invisible BS is better than the state of the visible BS if the distance of an invisible BS is shorter than the nearest visible BS. Based on the analysis result of channel states, a coverage probability and an average achievable rate are derived for 5G fractal small cell networks accounting for the proposed dual-directional path loss model. The main contributions of this paper are as follows:
\begin{enumerate}
\item A dual-directional path loss model incorporating both LoS and NLoS transmissions is proposed for 5G wireless fractal small cell networks.
\item Based on the proposed dual-directional path loss model, the coverage probability and the average achievable rate are derived for 5G wireless fractal small cell networks.
\item Numerical results indicate that the average achievable rate increases with the increase of the intensity of BSs and the decrease of the intensity of blockages, respectively. However, the minimum intensity of blockages and the maximum intensity of BSs  can not guarantee to achieve the maximum average achievable rate in 5G fractal small cell networks.
\end{enumerate}

The remainder of this paper is structured as follows. Section II describes the system model. The coverage probability and the average achievable rate are presented in Section III. The numerical results are discussed in Section IV. Finally, the conclusions are drawn in Section VII.

\section{System Model}
In this paper, BSs and rectangle blockages are assumed to be independently and randomly deployed in a plane. Moreover, the main assumptions of 5G fractal small cell networks are summarized as follows.
\begin{assumption}
BSs are assumed to be deployed in the infinite plane $\mathbb{R}^2$ according to a homogenous Poisson point process (PPP) ${\Phi _B}$ of intensity ${\lambda _B}$, which is denoted as \[{\Phi _B} = \left\{ {{x_i},i = 1,2,3,...} \right\},\tag{1}\] where $x_i$ is a two dimensional Cartesian coordinate, denoting the location of $B{S_i}$. $x_i$ is further extended as a Polar coordinate, i.e., $\left( {{r_i},{\theta _i}} \right)$, where ${r_i}$ is the distance between $B{S_i}$ and the coordinate origin, and satisfies ${r_1} < {r_2} < {r_3} <  \cdots  < {r_{i - 1}} < {r_i} <  \cdots $, ${\theta _i}$ is the coordinate azimuth of $B{S_i}$.
\end{assumption}
\begin{assumption}
Blockages are assumed to form a Boolean scheme of rectangles \cite{R5}. The centers of the rectangles form a homogeneous PPP ${\Phi _C}$ of intensity ${\lambda _C}$, which is denoted as \[{\Phi _C} = \left\{ {{y_j},j = 1,2,3,...} \right\},\,\tag{2}\]where ${y_j}$ denotes the location of the center of a rectangle in a two dimensional Cartesian coordinate. The length ${l_j}$ and width ${w_j}$ of a rectangle are independent each other. The coordinate azimuth of a rectangle is denoted as ${\varphi _j}$ and is uniformly distributed in $[0,2\pi )$. To simplify calculations, the length ${l_j}$, width ${w_j}$ and coordinate azimuth ${\varphi _j}$ are assumed to be constants $l$, $w$ and $\varphi$, respectively. Hence, in this paper the rectangle blockages can be modeled by marked Poisson point process ${\Phi _C} \times \left\{ {l,w,\varphi } \right\}$.
\end{assumption}
\begin{assumption}
To evaluate the received signal at a typical MU, in this paper the MU is assumed to be located at the coordinate origin. Channel models of BSs are assumed to be governed by independent identically distributed (i.i.d.) Rayleigh fading models. Furthermore, the received signal power ${P_i}$ at the typical MU from $B{S_i}$ can be expressed as \[{P_i} = {P_T}{L_i}{h_i},\,\tag{3}\]where ${h_i}$ denotes the Rayleigh fading between the typical MU and $B{S_i}$, which is distributed as an exponential distribution with the mean of 1\cite{R14}. For simplicity but without loss of generality, all BSs are assumed to transmit with the same transmission power ${P_T}$. ${L_i}$ denotes the path loss between the typical MU and $B{S_i}$. Moreover, the 5G wireless fractal small cell network is assumed to be interference limited and thermal noise is ignored. The system model of 5G wireless fractal small cell networks is illustrated as Fig. 1.
\begin{figure}[!t]
\centering
\includegraphics[width=9cm, height = 8cm, draft=false]{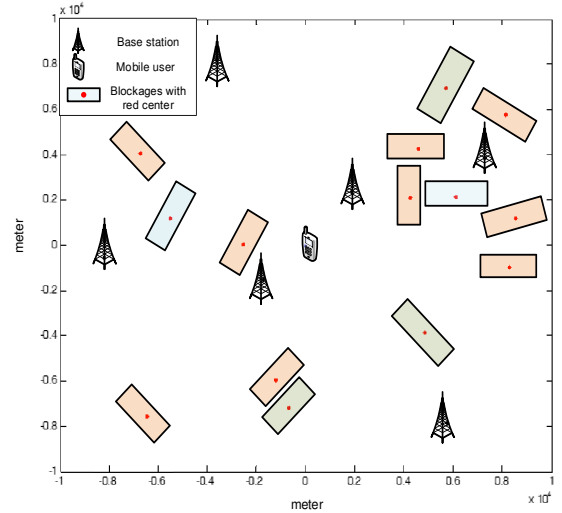}
\caption{The system model of 5G wireless fractal small cell networks.}
\label{fig1}
\end{figure}
\end{assumption}
\begin{assumption}
A MU is assumed to associate with the BS from which the MU receives the maximum signal power\cite{R13}. Without loss of generality, the BS associated with the typical MU is denoted as $B{S_k}$, where \[k = \mathop {\arg \max }\limits_{i \in {\Phi_B} } {P_i}.\,\tag{4}\]
\end{assumption}

 The path loss is associated with the distance and direction of the link relative to the BS. The wireless transmission between the typical MU and $B{S_i}$ is NLoS when blockages are existed in the link, whereas the wireless transmission is LoS. Thus, the dual-directional path loss ${L_i}$ considering differences between LoS and NLoS wireless transmissions is expressed as
\[\begin{array}{l}
{L_i}\left( {{x_i},{\Phi _C} \times \left\{ {l,w,\varphi } \right\}} \right)\\
 = \left\{ \begin{array}{l}
{r_i}^{ - {\alpha _{LoS}}},\quad with\;\Pr _i^{LoS}\left( {{x_i},{\Phi _C} \times \left\{ {l,w,\varphi } \right\}} \right)\\
{r_i}^{ - {\alpha _{NLoS}}},\quad with\;\Pr _i^{NLoS}\left( {{x_i},{\Phi _C} \times \left\{ {l,w,\varphi } \right\}}  \right)
\end{array} \right.,
\end{array}\,\tag{5}\]
where ${\alpha _{LoS}}$ and ${\alpha _{NLoS}}$ are LoS and NLoS wireless transmission path loss exponents, respectively. And two path loss exponents satisfy ${\alpha _{LoS}} < {\alpha _{NLoS}}$. $\Pr _i^{LoS}\left( {{x_i},{\Phi _C} \times \left\{ {l,w,\varphi } \right\}} \right) = \Pr _i^{LoS}\left( {{x_i}} \right)$ and $\Pr _i^{NLoS}\left( {{x_i},{\Phi _C} \times \left\{ {l,w,\varphi } \right\}} \right) = \Pr _i^{NLoS}\left( {{x_i}} \right)$ denote probabilities of LoS and NLoS wireless transmissions between the typical MU and $B{S_i}$, respectively. $\Pr \left\{  \cdot  \right\}$ is a probability operation.

\section{Coverage Probability and Average Achievable Rate}
In this section, the coverage probability and the average achievable rate of 5G wireless fractal small cell networks are derived based on dual-directional path loss model.
\subsection{Coverage Probability}
The coverage probability of the 5G wireless fractal small cell network is defined as the probability that the received signal-to-interference-ratio (SIR) at the typical MU is larger than a threshold value $T$. The general expression of SIR at the typical MU associated with $B{S_k}$ can be expressed as\[SI{R_k} = \frac{{{P_k}}}{{\sum\limits_{{\Phi _B,{i \ne k}}} {{P_i}} }} = \frac{{{L_k}{h_k}}}{{\sum\limits_{{\Phi _B,{i \ne k}}} {{L_i}{h_i}} }} = \frac{{{L_k}{h_k}}}{{{I_k}}},\,\tag{6}\]where ${P_i}$ and ${h_i}\left( {x_i \in \Phi_B ,i \ne k} \right)$ denote the received signal power and the channel fading from the interfering $B{S_i}$, respectively. ${L_k}$ is the path loss of the associated $B{S_k}$ and ${L_i}\left( i=1,2,...,k-1,k+1,...\right)$ is the path loss of the interfering $B{S_i}$. ${I_k}$ is the interference aggregated at the typical MU. Furthermore, when the typical MU associated with $B{S_k}$, the coverage probability of the 5G wireless fractal small cell network can be derived by
\[\begin{array}{l}
{p_k}\left( T \right) = \Pr \left\{ {SI{R_k} > T} \right\} = {{\mathbb{E}}_{{I_k},{L_k}}}\left[ {{e^{ - \frac{{T{I_k}}}{{{L_k}}}}}} \right]\\
 = {{\mathbb{E}}_{{L_k}}}\left[ {{\mathcal{L}_{{I_k}}}\left( {\frac{T}{{{L_k}}}} \right)} \right] = {{\mathbb{E}}_{{x_k},{\alpha _k}}}\left[ {{\mathcal{L}_{{I_k}}}\left( {\frac{T}{{{L_k}}}} \right)} \right]\\
 = \int\limits_0^\infty  {\frac{1}{{2\pi }}\left( {\int\limits_0^{2\pi } {{\mathcal{L}_{{I_k}}}\left( {\frac{T}{{{r_k}^{ - {\alpha _N}}}}} \right)\Pr _k^{NLoS}\left( {{x_k}} \right)d{\theta _k}} } \right.} \\
\left. {{\rm{ + }}\int\limits_0^{2\pi } {{\mathcal{L}_{{I_k}}}\left( {\frac{T}{{{r_k}^{ - {\alpha _L}}}}} \right)\Pr _k^{LoS}\left( {{x_k}} \right)d{\theta _k}} } \right){f_r}\left( {{r_k}} \right)dr,
\end{array}\,\tag{7}\]
where ${\mathcal{L}_{{I_k}}}\left( s \right)$ denotes the Laplace transform of ${I_k}$ evaluated at $s$, ${f_r}\left( {{r_k}} \right)$ is the probability density function (PDF) of the distance ${r_k}$ between the typical MU and $B{S_k}$. ${\mathbb{E}}\left[  \cdot  \right]$ is an expectation operation.

Before calculating the coverage probability of the 5G wireless fractal small cell network, probabilities of the LoS transmission and the NLoS transmission from $B{S_i}$ to the typical MU should be considered firstly. The transmission from $B{S_i}$ is LoS when no blockage is existed in the link between the typical MU and $B{S_i}$. Based on the Lemma 2 in \cite{R5}, the number of blockages $N$ in the link is a Poisson random variable with the mean of
\[{\mathbb{E}}\left[ N \right] = {\lambda _c}\left( {{r_i}l\left| {\sin \left( {\varphi  - {\theta _i}} \right)} \right| + {r_i}w\left| {\cos \left( {\varphi  - {\theta _i}} \right)} \right| + wl} \right).\,\tag{8}\]
Furthermore, the probability of the LoS transmission from $B{S_i}$ to the typical MU is derived by
\[\begin{array}{l}
\Pr _i^{LoS}\left( {{x_i}} \right) = \Pr \left\{ {N = 0} \right\}\\
 = {e^{ - {\lambda _c}\left( {{r_i}l\left| {\sin \left( {\varphi  - {\theta _i}} \right)} \right| + {r_i}w\left| {\cos \left( {\varphi  - {\theta _i}} \right)} \right| + wl} \right)}}.
\end{array}\,\tag{9}\]
Moreover, the probability of the NLoS transmission from $B{S_i}$ to the typical MU is derived by
\[\begin{array}{l}
\Pr _i^{NLoS}\left( {{x_i}} \right) = 1 - \Pr _i^{LoS}\left( {{x_i}} \right)\\
 = 1 - {e^{ - {\lambda _c}\left( {{r_i}l\left| {\sin \left( {\varphi  - {\theta _i}} \right)} \right| + {r_i}w\left| {\cos \left( {\varphi  - {\theta _i}} \right)} \right| + wl} \right)}}.
\end{array}\,\tag{10}\]

The interference ${I_k}$ aggregated at the typical MU associated with $B{S_k}$ is further expressed as
\[{I_k} = \sum\limits_{{\Phi _B}/{x_k}} {{L_i}{h_i}}  = \sum\limits_{{\Phi _B}/{x_k}} {{r_i}^{ - {\alpha _i}}{h_i}} .\tag{11}\]
Based on the dual-directional path loss model and probabilities of the LoS transmission and the NLoS transmission from $B{S_i}$ to the typical MU, it is straightforward to derive the Laplace transform ${\mathcal{L}_{{I_k}}}\left( s \right)$ of the aggregated interference ${I_k}$ as
\[\begin{array}{l}
{\mathcal{L}_{{I_k}}}\left( s \right) = {{\mathbb{E}}_{{I_k}}}\left[ {{e^{ - s{I_k}}}} \right]\\
 = {{\mathbb{E}}_{\Phi_B /{x_k},\{ {\alpha _i}\} ,\left\{ {{h_i}} \right\}}}\left[ {\exp \left( { - s\sum\limits_{\Phi_B /{x_k}} {{r_i}^{ - {\alpha _i}}{h_i}} } \right)} \right]\\
 = {{\mathbb{E}}_{\Phi_B /{x_k}}}\left[ {\prod\limits_{\Phi_B /{x_k}} {{{\mathbb{E}}_{\alpha ,h}}\left[ {\exp \left( { - s{r_i}^{ - \alpha }h} \right)} \right]} } \right]\\
 = {{\mathbb{E}}_{\Phi_B /{x_k}}}\left[ {\prod\limits_{\Phi_B /{x_k}} {\left[ {\frac{{{e^{ - {\lambda _c}\left( {{r_i}l\left| {\sin \left( {\varphi  - {\theta _i}} \right)} \right| + {r_i}w\left| {\cos \left( {\varphi  - {\theta _i}} \right)} \right| + wl} \right)}}}}{{s{r_i}^{ - {\alpha _L}} + 1}}} \right.} } \right.\\
\left. { + \left. {\frac{{1 - {e^{ - {\lambda _c}\left( {{r_i}l\left| {\sin \left( {\varphi  - {\theta _i}} \right)} \right| + {r_i}w\left| {\cos \left( {\varphi  - {\theta _i}} \right)} \right| + wl} \right)}}}}{{s{r_i}^{ - {\alpha _N}} + 1}}} \right]} \right]\\
\mathop  = \limits^{\left( a \right)} \exp \left( { - {\lambda _B}\int\limits_0^{2\pi } {\int\limits_{{r_k}}^\infty  {\left( {\frac{{s{r^{ - {\alpha _N}}}}}{{\left( {s{r^{ - {\alpha _N}}} + 1} \right)}} - \frac{{\left( {s{r^{ - {\alpha _N}}} - s{r^{ - {\alpha _L}}}} \right)}}{{\left( {s{r^{ - {\alpha _L}}} + 1} \right)}}} \right.} } } \right.\\
\left. {\left. { \times \frac{{{e^{ - {\lambda _c}\left( {rl\left| {\sin \left( {\varphi  - \theta } \right)} \right| + rw\left| {\cos \left( {\varphi  - \theta } \right)} \right| + wl} \right)}}}}{{\left( {s{r^{ - {\alpha _N}}} + 1} \right)}}} \right)rdrd\theta } \right)\\
\mathop  \le \limits^{(b)} \exp \left( { - 2\pi {\lambda _B}} \right.\int\limits_{{r_k}}^\infty  {\left( {\frac{{s{r^{ - {\alpha _N}}}}}{{\left( {s{r^{ - {\alpha _N}}} + 1} \right)}} - \frac{{\left( {s{r^{ - {\alpha _N}}} - s{r^{ - {\alpha _L}}}} \right)}}{{\left( {s{r^{ - {\alpha _L}}} + 1} \right)}}} \right.} \\
\left. {\left. { \times \frac{{{e^{ - {\lambda _c}wl}}{I_0}\left( {{\lambda _c}r\sqrt {{l^2} + {w^2}} } \right)}}{{\left( {s{r^{ - {\alpha _N}}} + 1} \right)}}} \right)rdr} \right),
\end{array}\,\tag{12}\]
where the step (a) of (12) is obtained by the probability generating functional (PGFL) of PPP \cite{R15} and like terms merger. The step (b) is the upper bound of the Laplace transform ${\mathcal{L}_{{I_k}}}\left( s \right)$ due to ${r_i}l\left| {\sin \left( {\varphi  - {\theta _i}} \right)} \right| + {r_i}w\left| {\cos \left( {\varphi  - {\theta _i}} \right)} \right| + wl \ge {r_i}l\sin \left( {\varphi  - {\theta _i}} \right) + {r_i}w\cos \left( {\varphi  - {\theta _i}} \right) + wl$. For simplicity but without loss of generality, $\varphi$ is assumed to be 0. It can be obtained that $\int\limits_0^{2\pi } {{e^{ - {\lambda _c}\left( {{r_i}l\sin \left( {{\theta _i}} \right) + {r_i}w\cos \left( {{\theta _i}} \right) + wl} \right)}}d\theta }  = 2\pi {e^{ - {\lambda _c}wl}}{I_0}\left( {{\lambda _c}r\sqrt {{l^2} + {w^2}} } \right)$, where ${I_0}\left(  \cdot  \right)$ is the modified Bessel function of the first kind. Submitting $s = \frac{T}{{r_k^{ - {\alpha _k}}}}$ into (12), the coverage probability of the 5G wireless fractal small cell network is derived by
\[\begin{array}{l}
{p_k}\left( T \right) = \int\limits_0^\infty  {\frac{1}{{2\pi }}} \left( {{\mathcal{L}_{{I_k}}}\left( {\frac{T}{{{r_k}^{ - {\alpha _N}}}}} \right)\left( {2\pi  - Q\left( {{r_k}} \right)} \right)} \right.\\
\left. {{\rm{               + }}{\mathcal{L}_{{I_k}}}\left( {\frac{T}{{{r_k}^{ - {\alpha _L}}}}} \right)Q\left( {{r_k}} \right)} \right){f_r}\left( {{r_k}} \right)d{r_k},
\end{array}\,\tag{13}\]
where $Q\left( {{r_k}} \right) = \int\limits_0^{2\pi } {{e^{ - {\lambda _c}\left( {{r_k}l\left| {\sin \left( {{\theta _k}} \right)} \right| + {r_k}w\left| {\cos \left( {{\theta _k}} \right)} \right| + wl} \right)}}d{\theta _k}} $, the PDF of ${f_r}\left( {{r_k}} \right)$ \cite{R16} is
\[{f_r}\left( {{r_k}} \right) = \frac{{2{{\left( {\pi {\lambda _B}} \right)}^k}}}{{(k - 1)!}}{r_k}^{2k - 1}{e^{ - \pi {\lambda _B}r_k^2}}.\,\tag{14}\]
\subsection{Average Achievable Rate}
The average achievable rate is another important performance metric in a wireless communication system. It gives a maximum data rate that a cellular network can support on the fixed bandwidth. The average achievable rate normalized by the bandwidth is expressed as\[{R_{avg}} = {\mathbb{E}}\left[ {{{\log }_2}\left( {1 + SIR} \right)} \right].\,\tag{15}\]

According to the Assumption 4, the typical MU is assumed to associate with the BS from which the typical MU can received the maximum receive signal power. The selected BS could be either the nearest $B{S_1}$, or other $B{S_i}\left( {i = 2,3, \ldots } \right)$ due to the fractal characteristics of 5G wireless small cell networks \cite{R17}. Thus, the average achievable rate of the 5G wireless fractal small cell network is expressed as\[\begin{array}{c}
{R_{avg}} = {{\mathbb{E}}_k}\left[ {{{\mathbb{E}}_{SI{R_k}}}\left[ {{{\log }_2}\left( {1 + SI{R_k}} \right)} \right]} \right]\\
 = \sum\limits_{k = 1}^\infty  {p_{ass}^k \times {{\mathbb{E}}_{SI{R_k}}}\left[ {{{\log }_2}\left( {1 + SI{R_k}} \right)} \right]},
\end{array}\,\tag{16}\]where $p_{ass}^k$ is the probability of the typical MU associated with $B{S_k}$. Due to the complexity of the user association scheme in the Assumption 4, the user association scheme is simplified by Table I. Based on the simplified user association scheme, the probability of the typical MU associated with $B{S_k}$ is derived by
\[\begin{array}{l}
p_{ass}^k = \int { \cdots \int {\left( {\frac{{\rm{1}}}{{{{\left( {{\rm{2}}\pi } \right)}^{k + 1}}}}\left( {\prod\limits_{i = 1}^{k - 1} {\Pr _i^{NLoS}\left( {{x_i}} \right)} } \right)} \right.} } \\
 \times \Pr _k^{LoS}\left( {{x_k}} \right)\Pr \left\{ {{r_1} > r_k^{{\alpha _L}/{\alpha _N}}} \right\}\\
 + \left( {\prod\limits_{i = 1}^k {\Pr _i^{NLoS}\left( {{x_i}} \right)} } \right)\Pr _{k + 1}^{NLoS}\left( {{x_{k + 1}}} \right)\\
\left. { + \left( {\prod\limits_{i = 1}^k {\Pr _i^{NLoS}\left( {{x_i}} \right)} } \right)\Pr _{k + 1}^{LoS}\left( {{x_{k + 1}}} \right)\Pr \left\{ {{r_k} < r_{k + 1}^{{\alpha _L}/{\alpha _N}}} \right\}} \right)\\
 \times f\left( {{r_1},{r_2},{r_3}, \cdots ,{r_k},{r_{k + 1}}} \right)d{r_1} \cdots d{r_{k + 1}}d{\theta _1} \cdots d{\theta _{k + 1}}
\end{array},\tag{17}\]
\emph{Proof}: See Appendix.
\begin{table}[!t]
\renewcommand{\arraystretch}{1.3}
\caption{The Simplified User Association Scheme.}
\label{tab1}
\centering
\begin{tabular}{l c r}
\hline
\textbf{Input}: Received signal powers of the typical MU, ignoring the channel \\fading.\\
\textbf{Output}: The selected BS.\\
For $i = 1\;to\;\infty $ do\\
\quad If ${r_1}^{ - {\alpha _1}} > {r_2}^{ - {\alpha _2}}$, the selected BS is the nearest BS,$B{S_1}$;\\
\quad Else ${r_2}^{ - {\alpha _2}} > {r_1}^{ - {\alpha _1}} \& {r_2}^{ - {\alpha _2}} > {r_3}^{ - {\alpha _3}}$, the selected BS is $B{S_2}$;\\
\quad\vdots\\
\quad Else ${r_k}^{ - {\alpha _k}} = \arg \max \left\{ {{r_{k + 1}}^{ - {\alpha _{k + 1}}},{r_k}^{ - {\alpha _k}}, \ldots ,{r_2}^{ - {\alpha _2}},{r_1}^{ - {\alpha _1}}} \right\}$\\
\quad \quad \quad (local maximum value), the selected BS is $B{S_k}$;\\
\quad End\\
End\\
As a result, the distance ${r_i}$ and the corresponding path loss exponent ${\alpha _i}$ \\are used to determine the selected BS. \\
\hline
\end{tabular}
\end{table}

Moreover, based on the definition of the coverage probability ${p_k}\left( T \right)$ in (7) which is the complementary cumulative distribution function (CCDF) of $SI{R_k}$, the PDF of $SI{R_k}$ is calculated by\[\begin{array}{l}
{f_{SI{R_k}}}\left( t \right) = \frac{{\partial \left( {1 - {p_k}\left( t \right)} \right)}}{{\partial t}}\\
 =  - \int\limits_0^\infty  {\frac{1}{{2\pi }}} \left( {} \right.\frac{{\partial {\mathcal{L}_{{I_k}}}\left( {\frac{t}{{{r_k}^{ - {\alpha _N}}}}} \right)}}{{\partial T}}\left( {2\pi  - Q\left( {{r_k}} \right)} \right)\\
 + \frac{{\partial {\mathcal{L}_{{I_k}}}\left( {\frac{t}{{{r_k}^{ - {\alpha _L}}}}} \right)}}{{\partial t}}Q\left( {{r_k}} \right)\left. {} \right){f_r}\left( {{r_k}} \right)d{r_k},
\end{array}\,\tag{18a}\]
with\[\frac{{\partial {\mathcal{L}_{{I_k}}}\left( {\frac{t}{{{r_k}^{ - {\alpha _N}}}}} \right)}}{{\partial t}}\mathop  = \limits^{s = \frac{t}{{{r_k}^{ - {\alpha _N}}}}} \frac{1}{{{r_k}^{ - {\alpha _N}}}}\frac{{\partial {\mathcal{L}_{{I_k}}}\left( s \right)}}{{\partial s}},\tag{18b}\]
\[\frac{{\partial {\mathcal{L}_{{I_k}}}\left( {\frac{t}{{{r_k}^{ - {\alpha _L}}}}} \right)}}{{\partial t}}\mathop  = \limits^{s = \frac{t}{{{r_k}^{ - {\alpha _L}}}}} \frac{1}{{{r_k}^{ - {\alpha _L}}}}\frac{{\partial {\mathcal{L}_{{I_k}}}\left( s \right)}}{{\partial s}},\,\tag{18c}\]
\[\begin{array}{l}
\frac{{\partial {\mathcal{L}_{{I_k}}}\left( s \right)}}{{\partial s}} = \exp \left( { - 2\pi {\lambda _B}\int\limits_{{r_k}}^\infty  {\left( {\frac{{s{r^{ - {\alpha _N}}}}}{{\left( {s{r^{ - {\alpha _N}}} + 1} \right)}}} \right.} } \right.\\
\left. {\left. { - \frac{{\left( {s{r^{ - {\alpha _N}}} - s{r^{ - {\alpha _L}}}} \right){e^{ - {\lambda _c}wl}}{I_0}\left( {{\lambda _c}r\sqrt {{l^2} + {w^2}} } \right)}}{{\left( {s{r^{ - {\alpha _L}}} + 1} \right)\left( {s{r^{ - {\alpha _N}}} + 1} \right)}}} \right)rdr} \right)\\
 \times \left( { - 2\pi {\lambda _B}\int\limits_{{r_k}}^\infty  {\frac{1}{{{{\left( {s{r^{ - {\alpha _N}}} + 1} \right)}^2}}} - \frac{{\left( {{s^2}{r^{ - {\alpha _N} - {\alpha _L}}} - 1} \right)}}{{{{\left( {s{r^{ - {\alpha _L}}} + 1} \right)}^2}}}} } \right.\\
\left. {\left. { \times \frac{{\left( {{r^{ - {\alpha _L}}} - {r^{ - {\alpha _N}}}} \right)}}{{{{\left( {s{r^{ - {\alpha _N}}} + 1} \right)}^2}}}{e^{ - {\lambda _c}wl}}{I_0}\left( {{\lambda _c}r\sqrt {{l^2} + {w^2}} } \right)} \right)rdr} \right).
\end{array}\,\tag{18d}\]

With the condition that the typical MU associates with $B{S_k}$, the conditional average achievable rate is expressed as
\[\begin{array}{l}
{{\mathbb{E}}_{SI{R_k}}}\left[ {{{\log }_2}\left( {1 + SI{R_k}} \right)} \right] = \int {{{\log }_2}\left( {1 + T} \right){f_{SI{R_k}}}\left( T \right)dT} \\
 =  - \int {{{\log }_2}} (1 + T)\int\limits_0^\infty  {\frac{1}{{2\pi }}} \left( {\frac{{\partial {\mathcal{L}_{{I_k}}}\left( {\frac{T}{{{r_k}^{ - {\alpha _N}}}}} \right)}}{{\partial T}}\left( {2\pi  - Q\left( {{r_k}} \right)} \right)} \right.\\
 + \left. {\frac{{\partial {\mathcal{L}_{{I_k}}}\left( {\frac{T}{{{r_k}^{ - {\alpha _L}}}}} \right)}}{{\partial T}}Q\left( {{r_k}} \right)} \right){f_r}\left( {{r_k}} \right)d{r_k}dT\\
 =  - \int\limits_0^\infty  {\frac{1}{{2\pi }}} \int {{{\log }_2}} (1 + T)\left( {\frac{{\partial {\mathcal{L}_{{I_k}}}\left( {\frac{T}{{{r_k}^{ - {\alpha _N}}}}} \right)}}{{\partial T}}\left( {2\pi  - Q\left( {{r_k}} \right)} \right)} \right.\\
 + \left. {\frac{{\partial {\mathcal{L}_{{I_k}}}\left( {\frac{T}{{{r_k}^{ - {\alpha _L}}}}} \right)}}{{\partial T}}Q\left( {{r_k}} \right)} \right)dT{f_r}\left( {{r_k}} \right)d{r_k}.
\end{array}\,\tag{19}\]
As a result, the average achievable rate of the 5G wireless fractal small cell network is obtained by submitting formulas (16), (18), and (19) into (15).

\section{Simulation Results And Discussions}
In this section, numerical results of the coverage probability and the average achievable rate of the 5G wireless fractal small cell network are analyzed. The default parameters are configured as: ${\lambda _B} = \frac{1}{{{{800}^2}\pi }}$,${\lambda _C} = 0.002$; $l = 15m$, $w = 10m$; ${\alpha _L} = 2$, ${\alpha _N} = 5$; $T=-5dB$.

\begin{figure}[!t]
\centering
\includegraphics[width=8cm, height = 7cm, draft=false]{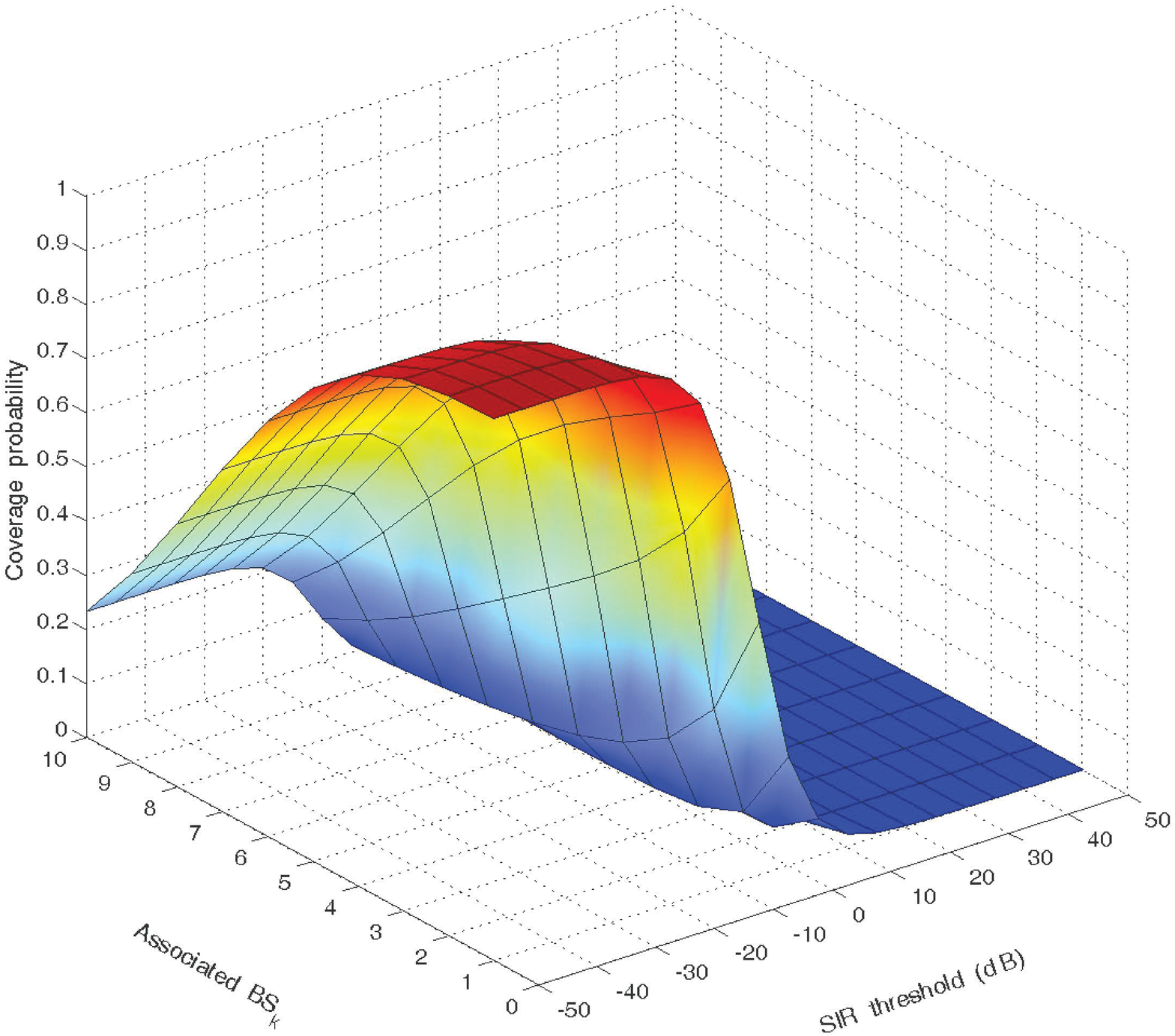}
\caption{The coverage probability of the 5G wireless fractal small cell network with respect to the associated $B{S_k}$ and the SIR threshold $T$.}
\label{fig2}
\end{figure}
Fig. 2 illustrates the coverage probability of the 5G wireless fractal small cell network with respect to the associated $B{S_k}$ and the SIR threshold $T$. When the SIR threshold $T$ is fixed, the coverage probability decreases with the increase of the value of $k$ marking the associated BS. The coverage probability decreases with the increase of the SIR threshold $T$ when the associated BS is fixed. Moreover, the coverage probability decreases more significantly when the value of $k$ is larger.

\begin{figure}[!t]
\centering
\includegraphics[width=8cm, height =7cm, draft=false]{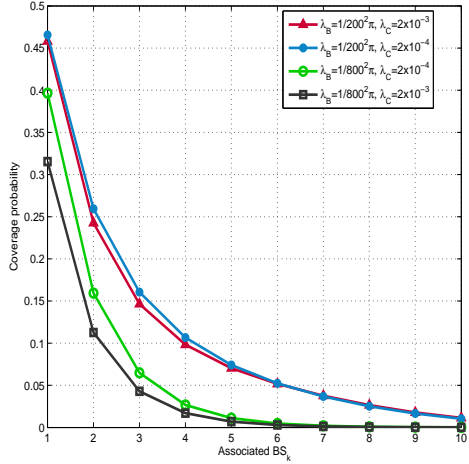}
\caption{The coverage probability of the 5G wireless fractal small cell network with respect to the associated $B{S_k}$ considering different intensities of BSs and blockages.}
\label{fig3}
\end{figure}
Fig. 3 shows the coverage probability of the 5G wireless fractal small cell network with respect to the associated $B{S_k}$ considering different intensities of BSs and blockages. When the intensity of BSs is fixed, the coverage probability decreases with the increase of the intensity of blockages from 0.0002 to 0.002. When the intensity of blockages is fixed, the coverage probability increases with the increase of the intensity of BSs from $\frac{1}{{{{800}^2}\pi }}$ to $\frac{1}{{{{200}^2}\pi }}$.

\begin{figure}[!t]
\centering
\includegraphics[width=8cm, height = 7cm, draft=false]{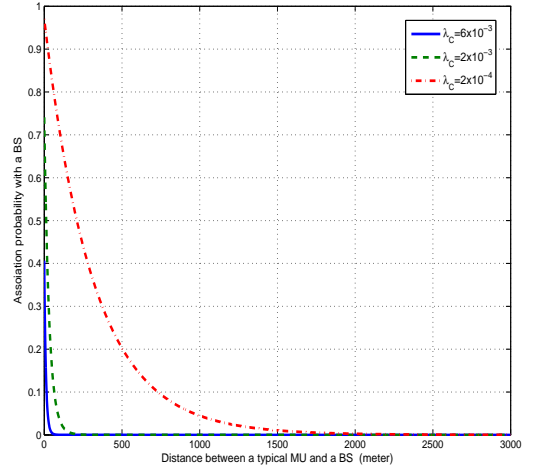}
\caption{The probability of LoS transmission with respect to the distance between a BS and a MU considering different intensities of blockages.}
\label{fig4}
\end{figure}
Fig. 4 illustrates the probability of LoS transmission with respect to the distance between a BS and a MU considering different intensities of blockages. The probability of the LoS transmission decreases with the increase of the distance between a BS and a MU. The curve of the probability of the LoS transmission declines more rapidly when the intensity of blockages is larger.

\begin{figure}[!t]
\centering
\includegraphics[width=8cm, height =7cm, draft=false]{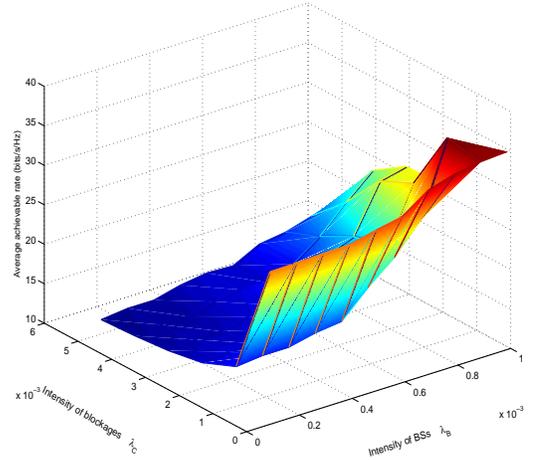}
\caption{The average achievable rate of the 5G wireless fractal small cell network with respect to intensities of BSs and blockages.}
\label{fig5}
\end{figure}
Fig. 5 illustrates the average achievable rate of the 5G wireless fractal small cell network with respect to intensities of BSs and blockages. The average achievable rate increases with the increase of the intensity of BSs when the intensity of blockages is fixed and large than 0.001. The average achievable rate decreases with the increase of the intensity of blockages when the intensity of BSs is fixed and smaller than 0.0008. The peak value comes out when the intensity of BSs and blockages are in the intervals $[0.0008, 0.001]$ and $[0.0001,0.001]$, respectively.\\

\section{Conclusions}
In this paper a dual-directional path loss model cooperating with LoS and NLoS transmissions is proposed for fractal small cell networks where the probability of LoS transmission is determined by the coordinate azimuth of the BS and the distance between a typical random MU and the BS. The coverage probability and the average achievable rate are analyzed based on the proposed path loss model. Numerical results indicate that the average achievable rate increases with increasing the intensity of BSs and blockages, respectively. Moreover, our results imply that the maximum average achievable rate is not achieved with the minimum intensity of blockages and the maximum intensity of BSs. Our results provide some guidelines for deployment of 5G fractal small cell networks.

\begin{appendix}
Based on the simplified user association scheme shown in Table I, the receive signal power of the selected BS has the maximum value. When the selected BS is the nearest $B{S_1}$, the conditional probability of the typical MU associated with $B{S_1}$ includes four items expressed as
\[\begin{array}{l}
p_{ass}^1\left( {{r_1},{r_2}} \right)\\
 = \Pr \left\{ {r_1^{ - {\alpha _N}} > r_2^{ - {\alpha _N}}|B{S_1}isNLOS,B{S_2}isNLOS} \right\}\\
 \times \Pr \left\{ {B{S_1}isNLOS,B{S_2}isNLOS} \right\}\\
 + \Pr \left\{ {r_1^{ - {\alpha _N}} > r_2^{ - {\alpha _L}}|B{S_1}isNLOS,B{S_2}isLOS} \right\}\\
 \times \Pr \left\{ {B{S_1}isNLOS,B{S_2}isLOS} \right\}\\
 + \Pr \left\{ {r_1^{ - {\alpha _L}} > r_2^{ - {\alpha _N}}|B{S_1}isLOS,B{S_2}isNLOS} \right\}\\
 \times \Pr \left\{ {B{S_1}isLOS,B{S_2}isNLOS} \right\}\\
 + \Pr \left\{ {r_1^{ - {\alpha _L}} > r_2^{ - {\alpha _L}}|B{S_1}isLOS,B{S_2}isLOS} \right\}\\
 \times \Pr \left\{ {B{S_1}isLOS,B{S_2}isLOS} \right\},
\end{array}\,\tag{20}\]
with ${r_2} > {r_1}$, and ${\alpha _N} > {\alpha _L}$. Furthermore, the probability $p_{ass}^1$ of the typical MU associated with $B{S_1}$ can be derived by
\[\begin{array}{l}
p_{ass}^1 = {\rm E}\left[ {p_{ass}^1\left( {{r_1},{r_2}} \right)} \right]\\
= \frac{1}{{4{\pi ^2}}}\int\limits_0^{2\pi } {\int\limits_0^{2\pi } {\int\limits_0^\infty  {\int\limits_0^{{r_2}} {p_{ass}^1\left( {{r_1},{r_2}} \right)f\left( {{r_1},{r_2}} \right)d{r_1}d{r_2}d{\theta _1}d{\theta _2}} } } } ,
\end{array}\,\tag{21a}\]
with
\[\begin{array}{l}
p_{ass}^1\left( {{r_1},{r_2}} \right)\\
 = \Pr _1^{NLoS}\left( {{x_1}} \right)\times \Pr _2^{NLoS}\left( {{x_2}} \right)\\
 + \Pr \left\{ {{r_1} < r_2^{{\alpha _L}/{\alpha _N}}} \right\}\Pr _1^{NLoS}\left( {{x_1}} \right)\times \Pr _2^{LoS}\left( {{x_2}} \right)\\
 + \Pr _1^{LoS}\left( {{x_1}} \right)
\end{array},\tag{21b}\]
where $f\left( {{r_1},{r_2}} \right)$ is the joint probability density function of ${r_1}$ and ${r_2}$. The entire expression of $p_{ass}^1$ is written as
\[\begin{array}{l}
{p_{ass}^1}
 = \frac{1}{{4{\pi ^2}}}\int\limits_0^{2\pi } {\int\limits_0^{2\pi } {\int\limits_0^\infty  {\int\limits_0^{{r_2}} {\Pr _1^{NLoS}\left( {{x_1}} \right)} } } } \\
 \times \Pr _2^{NLoS}\left( {{x_2}} \right)f\left( {{r_1},{r_2}} \right)d{r_1}d{r_2}d{\theta _1}d{\theta _2}\\
 + \frac{1}{{4{\pi ^2}}}\int\limits_0^{2\pi } {\int\limits_0^{2\pi } {\int\limits_0^\infty  {\int\limits_0^{r_2^{{\alpha _L}/{\alpha _N}}} {\Pr _1^{NLoS}\left( {{x_1}} \right)} } } } \\
 \times \Pr _2^{LoS}\left( {{x_2}} \right)f\left( {{r_1},{r_2}} \right)d{r_1}d{r_2}d{\theta _1}d{\theta _2}\\
 + \frac{1}{{2\pi }}\int\limits_0^{2\pi } {\int\limits_0^\infty  {\int\limits_0^{{r_2}} {\Pr _1^{LoS}\left( {{x_1}} \right)f\left( {{r_1},{r_2}} \right)d{r_1}d{r_2}d{\theta _1}} }}
\end{array}.\tag{22}\]

When the selected base station is $B{S_k}$, received signal powers of BSs satisfy $r_1^{ - {\alpha _1}} < r_2^{ - {\alpha _2}} <  \cdots  < r_k^{ - {\alpha _k}} > r_{k + 1}^{ - {\alpha _{k + 1}}}$. The probability of the typical MU associated with $B{S_k}$ is derived by
\[\begin{array}{l}
p_{ass}^k
= \frac{1}{{{{\left( {2\pi } \right)}^{k + 1}}}}\int \cdots{\int {p_{ass}^k\left( {{r_1}, \cdots ,{r_k},{r_{k + 1}}} \right)} } \\
\times f\left( {{r_1}, \cdots ,{r_k},{r_{k + 1}}} \right)d{r_1} \cdots d{r_{k + 1}}d{\theta _1} \cdots d{\theta _{k + 1}}
\end{array},\tag{23a}\]
with\[\begin{array}{l}
p_{ass}^k\left( {{r_1},{r_2},{r_3}, \cdots ,{r_k},{r_{k + 1}}} \right)\\
 = \left( {\prod\limits_{i = 1}^{k - 1} {\Pr _i^{NLoS}\left( {{x_i}} \right)} } \right) \Pr _k^{LoS}\left( {{x_k}} \right)\Pr \left\{ {{r_1} > r_k^{{\alpha _L}/{\alpha _N}}} \right\}\\
 + \left( {\prod\limits_{i = 1}^k {\Pr _i^{NLoS}\left( {{x_i}} \right)} } \right) \Pr _{k + 1}^{NLoS}\left( {{x_{k + 1}}} \right)\\
 + \left( {\prod\limits_{i = 1}^k {\Pr _i^{NLoS}\left( {{x_i}} \right)} } \right) \Pr _{k + 1}^{LoS}\left( {{x_{k + 1}}} \right)\Pr \left\{ {{r_k} < r_{k + 1}^{{\alpha _L}/{\alpha _N}}} \right\}.
\end{array}\,\tag{23b}\]
\end{appendix}

%

%

\end{document}